\documentclass[aps,prc,preprint]{revtex4}
\usepackage{amssymb}
\usepackage{graphics}
\usepackage{graphicx}
\begin{document}

\title{Representations in Density Dependent Hadronic Field Theory and compatibility with QCD
 sum-rules}

\author{R. Aguirre}
\email{aguirre@venus.fisica.unlp.edu.ar}
\affiliation{Departamento de F\'{\i}sica, Facultad de Ciencias Exactas,\\
Universidad Nacional de La Plata.\\
C. C. 67 (1900) La Plata, Republica Argentina.}

\date{\today}

\begin{abstract}
Different representations of an effective, covariant theory of the
hadronic interaction are examined. For this purpose we have
introduced nucleon-meson vertices parametrized in terms of scalar
combinations of hadronic fields, extending the conceptual frame of
the Density Dependent Hadronic Field Theory. Nuclear matter
properties at zero temperature are examined in the Mean Field
Approximation, including the equation of state, the Landau
parameters, and collective modes. The treatment of isospin
channels in terms of QCD sum rules inputs is outlined.
\end{abstract}


\maketitle

\section{Introduction}
A generalized concept in present theoretical physics is that
description of all physical phenomena should be derivable from
first principles in a unified way. However, the bridge towards
concrete applications requires elaborated procedures and judicious
arguments. This is the case of Quantum Chromodynamics (QCD), which
is the accepted theoretical model for the strong interactions.
Despite the fact that it is perturbative in the high energy
regime, the fundamental state of matter corresponds to the
opposite limit, where confinement and the breakdown of symmetries
make it mathematically intractable. Different effective models,
such as Nambu- Jona Lasinio, Skyrme, bag-like, and chiral
perturbation theory attempts to translate the main features of QCD
into the hadronic phase. On the other hand, lattice simulations
and
QCD sum-rules share this aim, although using different methods.\\
 QCD sum-rules is a ingenious procedure to
reveal the foundations of certain hadronic properties. The method
is based on the evaluation of correlation functions in terms of
quarks and gluons degrees, then applying the operator product
expansion it is possible to express it as combinations of
perturbative contributions and condensates (non-perturbative).
Finally this expansion is connected with the hadronic counterpart
by means of a Borel transformation.  The method was developed to
study meson \cite{SHIFMAN} as well as baryon
 \cite{IOFFE} properties in vacuum, it was subsequently generalized to
study finite density systems \cite{DRUKAREV0,DRUKAREV}. These
calculations provide a useful guide for some static properties of
hadrons immersed in a dense medium, as concerning QCD symmetries
and phenomenology. However, they are not able to take into account
the dynamical aspects of hadronic matter, which may be achieved by
inserting coherently these results into a theoretical model of
hadronic interactions.\\
A similar situation was found in the study of nuclear structure,
as the standard combination of \textit{microscopic} potentials and
the relativistic Dirac-Brueckner approximation for nuclear matter
gives rise to reliable, density dependent, nucleon self-energies.
Notwithstanding, this procedure was inadequate to treat finite
nuclei due to unsurmountable mathematical difficulties. A feasible
solution to this dilemma was proposed in
\cite{TOKI,HADDAD,LENSKE,TYPEL}, by defining density dependent
meson-nucleon vertices in terms of the self-energies obtained in
Dirac-Brueckner calculations. The enlarged hadronic model keeps
the mathematical versatility of the Quantum Hadrodynamic models
\cite{WALECKA}, but is equipped with couplings reflecting the
properties of the nuclear environment. The structure of spherical
nuclei, in the medium to heavy range, was studied within this
framework using the Hartree approximation.\\
 Further improvements,
developed in \cite{LENSKE}, replace the density dependence of the
meson-nucleon vertices by an expansion in terms of in-medium
nucleon condensates. This conceptual replacement restores the
covariance and the thermodynamical consistency of the model,
giving rise to the Density Dependent Hadronic Field Theory
(DDHFT). It was recently appended by introducing momentum
dependent vertices \cite{TYPEL2}, and a expansion of the vertices
in terms of meson mean field values \cite{AGUIRRE}.\\
The validity of the method is only justified \textit{a
posteriori}, and relies in the flexibility of hadronic field
models to accommodate pieces of information provided by another
field of research.  The enlarged model yields a simpler and more
intuitive description, instead of making involved calculations
based on first principles interactions. A significative
exemplification of this standpoint was given by the Brown-Rho
scaling of hadronic masses. Taking into account the chiral and
scale symmetries of QCD solely, an approximate scaling law for the
in-medium hadronic masses was derived in \cite{BROWNRHO}. This
hypothesis was applied to describe heavy ion collision, reaching
an excellent agreement with the experimental results for the low
mass dilepton production
rate \cite{BROWN}.\\
Given a set of physically meaningful self-energies as input for
DDHFT, there is room for a full family of hadronic models,
according to the field parameterization assumed for the vertices.
In principle, this may lead to different predictions for nuclear
observables. This point has not been investigated yet, and it is
the main purpose of the present work.\\
In a previous paper \cite{AGUIRRE}, the author attempted to make
use of the scheme outlined above to relate nuclear observables
with QCD inspired results. For this purpose we took as input the
nucleon self-energies in symmetric nuclear matter obtained in
\cite{DRUKAREV} by using QCD sum rules. However, Ref.
\cite{AGUIRRE} does not exhaust the physical description given
there, as it covers isospin asymmetric nuclear matter also. So, we
aim to complete our theoretical model including isospin degrees of
freedom, based on the results presented in \cite{DRUKAREV}.

We have organized this work presenting in the following section
the general features of DDHFT, and particularly the meson
parameterization of the hadronic vertices. In Section \ref{SecIII}
we present and discuss the results for symmetric nuclear matter. A
final summary is given in Section \ref{SecIV}.

\section{Density Dependent Hadronic Field Theory}\label{SecII}

As mentioned above, DDHFT was designed to incorporate into the
hadronic field formalism pieces of information produced by other
theoretical frames. Strictly speaking, DDHFT takes as input the
nucleon self-energies, which  can be decomposed into iso-scalar
and iso-vector components, each one containing Lorentz scalar and
vector contributions: \textit{i. e.} $\Sigma_s^{in}$,
$\Sigma_\mu^{in}$, $\Sigma_s^{in\;a}$, $\Sigma_\mu^{in\;a}$, where
the superscript $a$ distinguishes iso-vector quantities, and
$a=1,2$ stands for proton and neutron respectively.  For
homogeneous nuclear matter in steady state, they can be
parameterized as functions of continuous parameters, like baryonic
current density $j_\mu$ and temperature, characterizing the
macroscopic state of hadronic matter. In particular, there exists
a reference frame where the spatial part of the baryonic current
vanishes in the mean. In its former version $\Sigma^{in}$ was
taken from relativistic Brueckner-Hartree-Fock calculations with
one boson exchange potentials \cite{TOKI}.

On the other hand, a lagrangian density is proposed in terms of
meson fields $\sigma$, $\omega_\mu$ in the iso-scalar sector, and
$\phi_c$, $\rho_\mu^c$ in the iso-vector one. The indices
$c=1,2,3$ stand for the meson isospin projection
\begin{eqnarray}
{\mathcal L}&=& \bar{\Psi}(i \not \! \partial - M + \Gamma_s
\,\sigma- \Gamma_w \not \! \omega+\Gamma_f^c \,\phi_c+\Gamma_r^c
\, \not \! \rho_c)\, \Psi \nonumber
\\&+&\frac{1}{2}(\partial_\mu \sigma \partial^\mu \sigma\,-m_s^2
\sigma^2)-\frac{1}{4}W_{\mu
\nu} W^{\mu \nu} +\frac{1}{2} m_w^2 \omega_\mu \omega^\mu\\
\nonumber &+&\frac{1}{2}(\partial_\mu \phi_c
\partial^\mu \phi_c\,-m_f^2 \phi^2)-\frac{1}{4}R_{\mu \nu}^c R^{\mu \nu}_c+\frac{1}{2}
m_r^2 \rho_\mu^c \rho^\mu_c,\label{LAGRANGIAN}
\end{eqnarray}
where $W_{\mu \nu}=\partial_\mu \omega_\nu -
\partial_\nu \omega_\mu$, $R_{\mu \nu}^c=\partial_\mu \rho_\nu^c -
\partial_\nu \rho_\mu^c$ has been used, and $\Psi$ represents a isospinor.
We have adopted the convention that repeated isospin indices must
be summed. The isovector vertices $\Gamma^c$ take values over the
space generated by the Pauli matrices $\tau$ and the identity. The
masses have been fixed at the  phenomenological values $M=940$
MeV, $m_w=783$ MeV, $m_s=550$ MeV, $m_r=770$ MeV, and $m_f=984$
MeV. The iso-scalar vertices are assumed to depend on the scalar
combinations: $s_1=\sqrt{j_\mu j^\mu}$, $s_2=\bar{\Psi} \Psi$,
$m_1=\sqrt{\omega_\mu \omega^\mu}$, and $m_2=\sigma$, whereas the
isovector $\Gamma^c$ could depend linearly on
$j_c=\bar{\Psi}\tau^c\Psi$, or $\phi_c$, and on the scalars
$s_3=\sqrt{j_\mu^c j^\mu_c}$, $s_4=\sqrt{j_c j_c}$,
$m_3=\sqrt{\rho_\mu^c \rho^\mu_c}$, and $m_4=\sqrt{\phi_c
\phi_c}$. The baryonic and isospin current have been written as
$j_\mu=\bar{\Psi}\gamma_\mu \Psi$ and
$j_\mu^c=\bar{\Psi}\gamma_\mu \tau^c \Psi$, respectively. This is
not the most general dependence, but it keeps a clear separation
between isospin scalar and vector degrees of freedom. In Ref.
\cite{LENSKE} the dependence on $s_1$ or $s_2$ was presented and
 the case with only $s_2$ was explicitly studied; in \cite{AGUIRRE} the meson
dependence of the vertices was introduced within DDHFT, for
symmetric nuclear matter.

The field equations obtained for this lagrangian density are

\begin{eqnarray}
&&\left[ (i \not \! \partial - M + \Gamma_s \,\sigma- \Gamma_w
\not \! \omega) \delta_{AB} +\Gamma_{f\;AB}^c
\,\phi_c+\Gamma_{r\;AB}^c \, \not \! \rho_c
\right] \psi_B \nonumber\\
&&\;\;\;+ \bar{\Psi} \frac{\partial}{\partial \bar{\psi_A}} \left(
\Gamma_s \sigma - \Gamma_w\not \! \omega+ \Gamma_{f}^c
\,\phi_c+\Gamma_{r}^c \, \not \! \rho_c
\right) \Psi=0 \label{NUCLEONEQ}\\
&&\;\;\;\left( \square+m_s^2 \right)\sigma=\bar{\Psi}\left[
\Gamma_s+ \left(\frac{\partial \Gamma_s}{\partial m_2}
\sigma-\frac{\partial \Gamma_w}{\partial m_2} \not\! \omega\right) \right]
 \Psi \label{SIGMAEQ} \\
&&\;\;\;\partial_\mu W^{\mu \nu}+m_w^2 \omega^\nu=\bar{\Psi}\left[
\Gamma_w \gamma^\nu-\frac{\partial m_1}{\partial \omega_\nu}
\left(\frac{\partial \Gamma_s}{\partial m_1} \sigma-\frac{\partial
\Gamma_w}{\partial m_1} \not\!
\omega\right) \right]\Psi. \label{OMEGAEQ} \\
&&\;\;\;\left( \square+m_f^2 \right)\phi^c=\bar{\Psi}\left[
\Gamma_f^c+\frac{\partial m_4}{\partial \phi_c} \left(
\frac{\partial \Gamma_f^a}{\partial m_4}
\phi_a+\frac{\partial \Gamma_r^a}{\partial m_4} \not\! \rho_a\right) \right]
 \Psi \label{PHIEQ} \\
&&\;\;\;\partial_\mu R^{\mu \nu}_c+m_r^2
\rho^\nu_c=-\bar{\Psi}\left[ \Gamma_{r\,c}
\gamma^\nu+\frac{\partial m_3}{\partial \rho_\nu^c}
\left(\frac{\partial\Gamma_f^a}{\partial m_3}
\phi_a+\frac{\partial \Gamma_r^a}{\partial m_3} \not\!
\rho_a\right) \right]\Psi, \label{RHOEQ}
\end{eqnarray}
where the isospin indices have been denoted as: $A,B=1,2$ for
nucleons, and $a,c=1,2,3$ for mesons. In the first equation, the
vertex derivatives must be interpreted as
\begin{eqnarray}
 \frac{\partial \Gamma}{\partial \bar{\psi_A}}=\sum_1^4 \frac{\partial \Gamma}
 {\partial s_k}\frac{\partial s_k}{\partial \bar{\psi_A}} \nonumber.
\end{eqnarray}

The mean field approximation (MFA) is suited to describe
homogeneous matter. Within this scheme, meson fields are replaced
by their uniform mean values and bilinear combinations of spinors
are replaced by their expectation values. For homogeneous,
isotropic nuclear matter in steady state, some simplifications
arise in the meson mean values. For instance, their coordinate
dependence can be neglected; the spatial components of the vectors
$\omega_\mu$ and $\rho_\mu$ become null, and only the third
component of the isovector mesons are non-zero, due to isospin
conservation.\\
Therefore within MFA, Eqs. (\ref{NUCLEONEQ}-\ref{RHOEQ}) reduce to
\begin{eqnarray}
0=\left[(i\not\!\partial-M)\,\delta_{AB}+\Sigma_s^{AB}-\not \! \!
\Sigma^{AB}\right] \psi_B,&& \; \label{NUCLEONEQ2}\\
m_s^2 \tilde{\sigma}=<\bar{\Psi}\,\Gamma_s\Psi>
+<\bar{\Psi}\frac{\partial \Gamma_s}{\partial m_2} \Psi \,\sigma>
&-&<\bar{\Psi}\frac{\partial \Gamma_w}{\partial
m_2}\gamma_\mu \Psi \omega^\mu>; \label{SIGMAEQ2} \\
g^\nu_0\;m_w^2 \tilde{\omega}=<\bar{\Psi}\,\Gamma_w
\gamma^\nu\Psi>-<\bar{\Psi}\,\frac{\partial \Gamma_s} {\partial
m_1}\Psi\,\frac{\sigma\omega^\nu}{m_1} >
&+&<\bar{\Psi}\frac{\partial \Gamma_w}{\partial
m_1}\gamma_\mu\Psi \,\frac{\omega^\nu\omega^\mu}{m_1}>; \label{OMEGAEQ2}\\
 \delta_{3c}\;m_f^2 \tilde{\phi}=<\bar{\Psi}\Gamma_f^c
\Psi>+<\bar{\Psi}\frac{\partial \Gamma_f^a}{\partial
m_4}\Psi\,\frac{\phi_c\phi_a}{m_4}>&+&<\bar{\Psi}\frac{\partial
\Gamma_r^a}{\partial m_4}\gamma_\mu\Psi
\frac{\phi_c}{m_4} \rho_a^\mu>;  \label{PHIEQ2} \\
-\delta_{3c}g^\nu_0 m_r^2 \tilde{\rho}=<\bar{\Psi}\Gamma_{r\,c}
\gamma^\nu \Psi>+<\bar{\Psi} \frac{\partial\Gamma_f^a}{\partial
m_3}\Psi \frac{\rho^\nu_c\phi_a}{m_3}
>&+&<\bar{\Psi}\frac{\partial \Gamma_r^a}{\partial m_3}\gamma_\mu\Psi
\frac{\rho^\nu_c \rho^\mu_a}{m_3}>;\label{RHOEQ2}
\end{eqnarray}

where tildes over the meson symbols stand for their mean field
values, and the two last terms within brackets in Eq.
(\ref{NUCLEONEQ2}) can be regarded as the scalar and vector
components of the nucleon self-energies:
\begin{eqnarray}
\Sigma_s^{AB}&=&<\Gamma_s\sigma+\bar{\Psi}\Lambda_2\Psi>\delta_{AB}
\,+<\Gamma_f^{c\,AB}\phi_c+\tau_a^{AB}\,\bar{\Psi} \Lambda_4
\Psi\frac{\bar{\Psi}\tau_a\Psi}{s_4}> ,\\
\Sigma_\nu^{AB}&=&<\Gamma_w\omega_\nu-\bar{\Psi}\Lambda_1
\frac{j_\nu}{s_1} \Psi>\delta_{AB}
\,-<\Gamma_{r\,c}^{AB}\rho_\nu^c+\tau_a^{AB}\,\bar{\Psi}\Lambda_3
\Psi\frac{j_\nu^a}{s_3} >,
 \end{eqnarray}
with $\Lambda_k=\partial\left(\Gamma_s\sigma-\Gamma_w \not \!
\omega+\Gamma_f^c \phi_c+ \Gamma_r^c \not \!
\rho_c\right)/\partial s_k$ .

For the sake of concreteness, we have examined three possible
parameterizations:
\renewcommand\theenumi{\alph{enumi}}
\renewcommand\labelenumi{\theenumi)}
 \begin{enumerate}
 \item symmetric nuclear matter, with $\Gamma_s(s_1)$, $\Gamma_w(s_1)$;
 \item symmetric nuclear matter, with $\Gamma_s(s_2)$, $\Gamma_w(s_1)$;
 \item asymmetric nuclear matter, with $\Gamma_s(m_2)$,
 $\Gamma_w(m_1)$, $\Gamma_f^c(\phi^c,m_4)$,
 $\Gamma_r^c=\Gamma_r(\phi^c,m_3)$.
 \end{enumerate}
 The first two cases are comparable to the instances presented in
 \cite{LENSKE} as vector and scalar dependencies. The last case
 is a development of the preliminaries calculations of \cite{AGUIRRE},
 the field dependence of the isospin vertices have been chosen so
 as to adjust the QCD sum rules calculations of \cite{DRUKAREV}.\\
 These results are
simplified within the assumptions (a-c) above, for the sake of
completeness we consider separately each of these cases:
\begin{eqnarray}
\mbox{a)}&& \,\Sigma_s=\Gamma_s \tilde{\sigma},\,
\Sigma_\nu=g_{\nu 0}\left[
\Gamma_w\tilde{\omega}+\frac{d\Gamma_w}{dn_B}\,\tilde{\omega}n_B
-\frac{d\Gamma_s}{dn_B}\,\tilde{\sigma} n_s \right],\label{CASEA} \\
&&\tilde{\sigma}=\Gamma_s n_s/m_s^2, \, \tilde{\omega}=\Gamma_w
n_B/m_v^2, \nonumber \\
\mbox{b)}&&\, \Sigma_s=\tilde{\sigma}\left(\Gamma_s+
\frac{d\Gamma_s}{dn_s} n_s\right),\, \Sigma_\nu=g_{\nu
0}\,\tilde{\omega}
\left( \Gamma_w+\frac{d\Gamma_w}{dn_B}\,n_B\right), \label{CASEB}\\
&& \,\tilde{\sigma}, \,\mbox{and} \;\tilde{\omega}\, \mbox{have
the same expressions as in (a)}
,\nonumber\\
\mbox{c)}&&\,
\Sigma_s^{AB}=\Sigma_s^{isos}\delta_{AB}+\Sigma_s^{isov\,AB}, \;
\Sigma_\nu^{AB}=g_{\nu
0}\left(\Sigma_v^{isos}\delta_{AB}+\Sigma_v^{isov\,AB}\right),
\label{CASEC} \\
&&\Sigma_s^{isos}=
\Gamma_s\tilde{\sigma},\;\Sigma_s^{isov\,AB}=\Gamma^{3\,AB}_f
\tilde{\phi}, \;\Sigma_v^{isos}=\Gamma_w \tilde{\omega},
\;\Sigma_v^{isov\,AB}=-\Gamma_r^{3\,AB}\tilde{\rho},\nonumber\\
&&m_s^2 \tilde{\sigma}=\frac{d\Sigma_s^{isos}}{d\tilde{\sigma}}n_s
, \; m_w^2 \tilde{\omega}=
\frac{d\Sigma_v^{isos}}{d\tilde{\omega}}n_B,\nonumber\\
&&m_f^2\tilde{\phi}=
<\bar{\Psi}\frac{d\Sigma_s^{isov}}{d\tilde{\phi}}\Psi>-
<\bar{\Psi}\frac{d\Sigma_v^{isov}}{d\tilde{\phi}}\gamma_0\Psi>, \;
m_r^2\tilde{\rho}=<\bar{\Psi}\frac{d\Sigma_v^{isov}}{d\tilde{\rho}}\gamma_0\Psi>.\nonumber
\end{eqnarray}
 Obtaining the equations listed above, we have used the Wick theorem
and the Hartree approximation, therefore contractions of fields
belonging to different full contracted terms (both in Lorentz and
isospin indices) have been neglected. This approach allowed us to
extract the vertices $\Gamma$ and its derivatives from the
expectation values, and to express the mean value of products of
more than two fermion fields as product of baryonic densities and
currents. Finally, we have adopted the reference frame of static
matter, therefore we have $<j_\mu>=n_B\,g_{\mu 0}$,
$<j_\mu^c>=(n_B^{(1)}-n_B^{(2)})\,g_{\mu 0}\delta_{3c}$, with
$n_B^{A}=<\bar{\psi}_A\gamma_0 \psi_A>$ the number of neutrons or
protons per unit volume, for $A=2,1$ respectively, and
$n_B=n_B^{(1)}+n_B^{(2)}$ the baryonic number density. Furthermore
we have introduced the simplifying notation
$n_s^{(A)}=<\bar{\psi}_A\psi_A>$, $n_s=<\bar{\Psi}\Psi>$,
$\Gamma=<\Gamma>$, $s_1=n_B$, $s_2=n_s$, $s_3=\mid
n_B^{(1)}-n_B^{(2)}\mid$, $s_4=\mid n_s^{(1)}-n_s^{(2)}\mid$,
$m_1=\tilde{\omega}$, $m_2=\tilde{\sigma}$, $m_3=\tilde{\rho}$,
and $m_4=\tilde{\phi}$.

Cases (a) and (b) yield meson equations resembling those of the
Walecka model \cite{WALECKA}, and nucleon self-energies with the
\emph{rearrangement} contributions added. In the instance (c) the
self-energies have the same structure as in QHD-I, but the source
term of the meson equations are modified. With respect to the
isovector meson field equations, it must be noted that
$<\bar{\Psi}\tau_c \Psi>=0$ for $c=1,2$ as isospin is a conserved
charge. In order to give explicit expressions for the
corresponding vertices,
 we follow the parameterization of \cite{DRUKAREV}:
\begin{eqnarray}
\Sigma_k^{AA\,(in)}= \frac{\alpha_k +I_3^A \beta_k t}{\lambda
+n_0/n_B+I_3^A \,\alpha t},\;\;k=s,v\label{QCDSelf}
\end{eqnarray}
with $I_3^A=1 (-1)$, for protons (neutrons), $n_0=0.17 fm^{-3}$,
$t=(n_B^{(2)}-n_B^{(1)})/n_B$, the remaining coefficients can be
expressed in terms of the quantities used in Ref. \cite{DRUKAREV}
as $\lambda=C_v^q v_N+C_g^q g_N+C_{u1}^q+C_\omega^q A_{4q}^{q0}$,
$\alpha=C_{u2}^q+C_\omega^q A_{4q}^{q1}$, $\alpha_s=M\lambda-C_k^I
\kappa_N-C_\omega^I A_{4q}^{I0}$, $\beta_s=M \alpha-(C_\zeta^I
\zeta_N+C_\omega^I A_{4q}^{I1})$, $\alpha_v=-(C_v^p v_N+M
C_{u1}^p+M C_\omega^p A_{4q}^{p0}$), $\beta_v=-(C_{v-}^pv_N^-+M
C_{u2}^p+MC_\omega^p A_{4q}^{p0})$, which were obtained in
\cite{DRUKAREV}, by averaging over the Borel mass. For the sake of
completeness we give here the numerical values $v_N=3$, $v_N^-=1$,
$g_N=-8M/9$, $\kappa_N=8$, $\zeta_N=0.54$, $C_\kappa^I=-0.042$
GeV; $C_\zeta^I=-0.042$ GeV; $C_g^q=0.011/GeV$; $C_{v-}^p=-0.068$
GeV; $C_\omega^k=-0.067, -0.095, -0.070$ GeV,
$A_{4q}^{0k}=1.90,-0.57, -0.11$, and $A_{4q}^{1k}=-0.92, 0.09,
-0.21$ for $k=I, p$ and $q$ respectively; $C_v^k=-0.062,-0.090$,
$C_{u1}^k=-0.074, 0.094$, and
$C_{u2}^k=0.008, -0.02$ for $k=q$ and $p$ respectively. \\
From Eq. (\ref{QCDSelf}), we can extract isoscalar and isovector
contributions, \textit{i.e.} we consider the splitting
$\Sigma_k^{AA\,(in)}=\Sigma_k^{isos\,(in)}+\Sigma_{k\,A}^{isov\,(in)}$,
with
\begin{eqnarray}
\Sigma_k^{isos \,(in)}=\frac{\alpha_k n_B}{n_0+\lambda n_B},
\;\;\Sigma_{k\,A}^{isov\, (in)}=I_3^A
t\frac{\beta_k-\alpha'_k\alpha n_B}{\lambda+n_0/n_B+I_3^A \alpha
t},\label{QCDSRself}
\end{eqnarray}
with $\alpha'_k=\alpha_k/(n_0+\lambda n_B)$.\\
 A choice of the isovector interaction in Eq.
(\ref{LAGRANGIAN}), which is coherent with these results is
\begin{eqnarray}
\mathcal{L}_{iso}=\bar{\Psi}\tau^c\frac{G_r \not \! \rho_c+G_f
\phi_c}{1+g \,\tau_a \phi^a}\Psi,\nonumber
\end{eqnarray}
where $G_r$ is assumed to depend only on $m_3$, whereas $G_f$ and
$g$ are considered as functions of $m_4$ only. The following
vertices are deduced from it
\begin{eqnarray}
\Gamma_r^c=\frac{G_r \tau^c(1-g\tau_a\phi^a)}{1-g^2\phi^2},\;\;
\Gamma_f^c=\frac{G_f\tau^c(1-g\tau_a\phi^a)}{1-g^2\phi^2}.\nonumber
\end{eqnarray}
In the MFA, they give rise to the following contributions to the
nucleon self-energies
\begin{eqnarray}
\Sigma_v^{isov\,AB}=-\delta_{AB}\frac{I_3^A G_r
\tilde{\rho}}{1+I_3^A\,g \tilde{\phi}},\;\;
\Sigma_s^{isov\,AB}=\delta_{AB}\frac{I_3^AG_f
\tilde{\phi}}{1+I_3^A\,g \tilde{\phi}}.\nonumber
\end{eqnarray}
The isovector meson field equations can now be written as
\begin{eqnarray}
m_f^2\tilde{\phi}&=&\sum_{A=1}^2
\left(\frac{\partial\Sigma_s^{isov\,AA}}{\partial\tilde{\phi}}\,n_s^A-
\frac{\partial\Sigma_v^{isov\,AA}}{\partial\tilde{\phi}}\,n_B^A
\right), \;\;
m_r^2\tilde{\rho}=\sum_{A=1}^2\frac{\partial\Sigma_v^{isov\,AA}}{\partial\tilde{\rho}}
\,n_B^A. \nonumber
\end{eqnarray}

In DDHFT \cite{TOKI,HADDAD,LENSKE}, the isoscalar vertices
$\Gamma_s$ and $\Gamma_w$ are defined by means of the relations
\begin{equation}\Sigma_s^{(in)}=\Gamma_s n_s/m_s^2, \;\;\;
\Sigma_\mu^{(in)}=\Gamma_w n_\mu/m_w^2, \label{DDHFT}
\end{equation}
where $\Sigma_s^{(in)},\Sigma_\mu^{(in)}$ are the self-energies
obtained with one boson exchange potentials in the
Dirac-Brueckner-Hartree-Fock approach for symmetric nuclear
matter. It must be noted that the right hand sides of these
equations do not coincide in general with the dynamical
self-energies. Therefore Eqs. (\ref{DDHFT}) and
(\ref{CASEA})-(\ref{CASEB}) do not coincide, unless the
\emph{rearrangement} terms were omitted. In \cite{AGUIRRE}, the
author proposed that the vertices are solutions of someone of the
differential equations (\ref{CASEA})-(\ref{CASEC}), instead. They
must be solved together with the self-consistent condition for the
meson fields. The explicit form of part of these solutions have
been anticipated in \cite{AGUIRRE}, so we summarize them below and
add new results to the case (c) concerning isospin asymmetric
nuclear matter.
\begin{eqnarray}
\mbox{a)}\,\Gamma_s&=&\frac{\Sigma_s^{(in)}}{\tilde{\sigma}}, \;\;
\Gamma_w^2=2\,\left(\frac{ m_w}{n}\right)^2\int_0^{n_B} dn'
\left[\Sigma_v^{(in)}+\left(\frac{n_s}{m_s}\right)^2\Gamma_s\frac{d\Gamma_s}{dn'}
\right];\nonumber\\
\mbox{b)}\,\Gamma_s^2&=&2\,
\left(\frac{m_s}{n_s}\right)^2\int_0^{n_B} \,
dn'\frac{dn_s}{dn'}\,\Sigma_s^{(in)},\;\;\Gamma_w^2=2\,
\left(\frac{m_w}{n_B}\right)^2\int_0^{n_B} \, dn'\,\Sigma_v^{(in)};\nonumber \\
\mbox{c)}\,\Gamma_s&=&\Sigma_s^{isos
\,(in)}/\tilde{\sigma},\;\Gamma_w=\Sigma_v^{isos\,(in)}/\tilde{\omega},
\;g=\frac{\alpha t n_B}{\tilde{\phi}(n_0+\lambda n_B)},\nonumber\\
G_f&=&\frac{\beta_s-\alpha'_s \alpha n_B}{n_0+\lambda
n_B}\,\frac{t n_B}{\tilde{\phi}}, \, G_r=-\frac{\beta_v-\alpha'_v
\alpha n_B}{n_0+\lambda n_B}\,\frac{t n_B}{\tilde{\rho}},\nonumber \\
 \tilde{\sigma}^2&=&\frac{2}{m_s^2}\int_0^{n_B} \, dn'\,n_s\,
\frac{d\Sigma_s}{dn'}^{isos\,(in)},\;\;\tilde{\omega}^2=\frac{2}{m_w^2}\int_0^{n_B}
\, dn'\,n' \,\frac{d\Sigma_v}{dn'}^{isos\,(in)},\nonumber\\
\tilde{\phi}^2&=&\frac{1}{m_f^2}\sum_{A,C=1}^2(1-I_3^C
t)\int_0^{n_B} \, dn'\left[n_s^A
\frac{\partial\Sigma_{s\,A}}{\partial
n_C'}^{isov\,(in)}-\frac{1}{2}n'(1-I_3^A t)
\frac{\partial\Sigma_{v\,A}}{\partial
n_C'}^{isov\,(in)}\right],\nonumber \\
\tilde{\rho}^2&=&\frac{1}{2m_r^2}\sum_{A,C=1}^2(1-I_3^C t)(1-I_3^A
t)\int_0^{n_B} \, dn' n' \frac{\partial\Sigma_{v\,A}}{\partial
n_C'}^{isov\,(in)}.\nonumber
\end{eqnarray}

We have assumed that the input functions have been parameterized
in terms of the partial nucleon densities $n_C$, furthermore
within the two last eqs. of case (c), the isospin parameter $t$ is
held
constant in the integration.  \\
Thus we have obtained a set of relations, defining a hadronic
field model suited to reproduce the nucleon self-energies provided
by other theoretical framework. In particular, the meson dependent
vertices have been extended to deal with isospin asymmetric
matter. The resulting vertices are given as functions of $n_B$,
and eventually $s_3$, but they can be rewritten in terms of
the meson fields or nucleon condensates.\\
In the next section, these results will be compared with the
standard DDHFT treatment, and the ability to adjust the nuclear
matter phenomenology will be examined.

\section{Results and Discussion}\label{SecIII}

 A well known feature of isospin symmetric nuclear matter is its energy per
particle, having a minimum at a baryonic density about
$n_0=0.16\,fm^{-3}$. This property gives rise to bound states at
zero temperature,  and it is the least requirement that a model of
nuclear matter should satisfy. Starting with Eq.
(\ref{LAGRANGIAN}), we have evaluated the energy-momentum tensor
$T^{\mu \nu}$ by the canonical procedure, and the energy per unit
volume in the MFA is obtained by taking the in-medium expectation
value of $T^{00}$.

\begin{eqnarray}
E_{MFA}&=&\int_0^{p_F} \frac{d^3p}{(2 \pi)^3}\sqrt{p^2+M^{*
2}}+\Gamma_w n_B \tilde{\omega}
+n_s(\Sigma_s^{isos}-\Gamma_s\tilde{\sigma})+
\frac{1}{2}(m_s^2\tilde{\sigma}^2-m_w^2\tilde{\omega}^2),
\label{EOS}
\end{eqnarray}
where we have introduced the Fermi momentum $p_F$, related to the
baryonic density by $n_B=2p_F^3/(3 \pi^2)$, and the effective
nucleon mass $M^*=M-\Sigma_s$. The mean value of the isovector
meson fields becomes zero for symmetric nuclear
matter.\\
Finally, the binding energy is defined as $E_B=E_{MFA}/n_B-M$,
which should have a minimum value $E_B\simeq -16$ MeV at the
normal density $n_0$, to satisfy the nuclear matter phenomenology.
Using the thermodynamical relation $P=\mu n_B-E_{MFA}$ we have
obtained the pressure for the nuclear matter, with the chemical
potential given by $\mu=E_F+\Sigma_v^{isos}$, being
$E_F=\sqrt{p_F^2+M^{*2}}$.\\

\begin{figure}\vspace{-3cm}
\includegraphics[width=\textwidth]{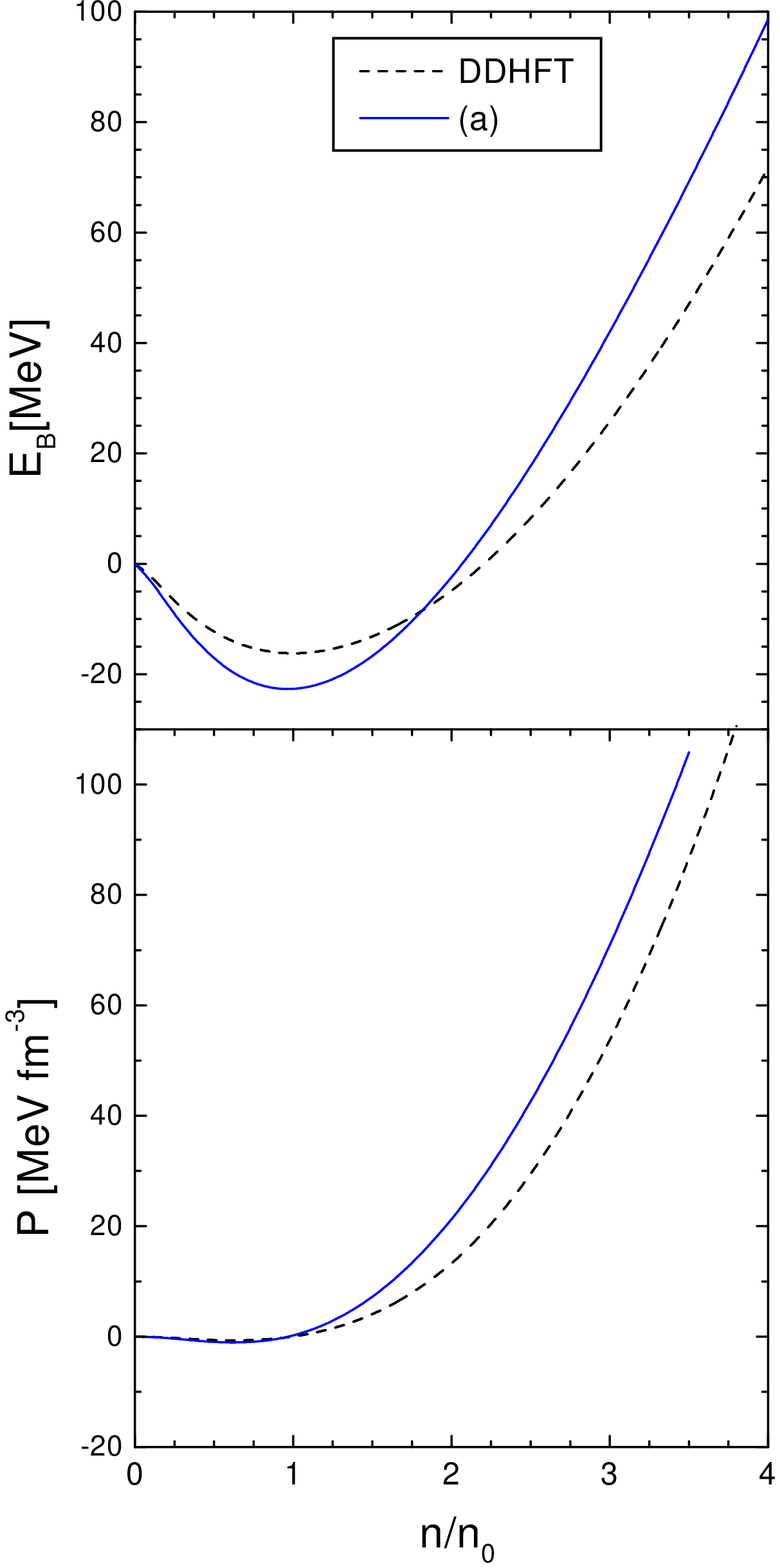}
\caption{The binding energy (top) and pressure (bottom) for
symmetric nuclear matter at zero temperature, evaluated in the MFA
using the standard DDHFT treatment and within the approach (a)
.\label{FIG1}}
\end{figure}

In first place, we have examined the effects of introducing the
effective vertices by means of the algebraic Eqs. (\ref{DDHFT}),
or by solving a differential equation of the type listed in Eqs.
(\ref{CASEA})-(\ref{CASEC}). It must be noted that the first
option has the property of reproducing the energy density of the
original Dirac-Brueckner calculations, whenever the
parameterization of the vertices in terms of only the baryon
number density is chosen. Otherwise, it is a definition without
any special physical significance. The alternative procedure
proposed in \cite{AGUIRRE}, aims to make a connection between the
hadronic vertices and another theoretical framework, by imposing
the equality of the nucleon self-energies evaluated in both cases.\\
 For the purpose stated, we have chosen as input the results of
 \cite{TYPEL}. In this work, an ansatz for the  vertices is proposed
 that fits several specific Dirac-Brueckner outcomes, but avoiding the
 unphysical behavior they exhibit in the zero density limit. The
 ansatze for the couplings $\Gamma_s^{TW}$ and $\Gamma_w^{TW}$ are rational
 functions of the relative baryonic density $n_B/n_0$. The
 corresponding self-energies are obtained as $\Sigma_s^{(in)}=\Gamma_s^{TW}
 \tilde{\sigma}$, $\Sigma_v^{(in)}=\Gamma_w^{TW} \tilde{\omega}$, with
 the meson mean field values given by equations similar to those
 of Eq. (\ref{CASEA}). For details, see Ref. \cite{TYPEL}.\\
 In Fig. \ref{FIG1} we show the binding energy and pressure evaluated
 within the standard DDHFT treatment of \cite{TYPEL}. Given the functions $\Sigma_s^{(in)},
 \,\Sigma_v^{(in)}$ as above, we can also use them to deduce the vertices
 $\Gamma_s,\,\Gamma_w$ which reproduce these self-energies within the approach (a);
\textit{i.e.} solving Eqs. (\ref{CASEA}). The corresponding
equation of state is shown in the same figure. We have examined
densities up to four times $n_0$, a range that may be reached
inside neutron stars. It can be seen that similar results for the
pressure are obtained for densities below $1.25 n_0$, but the
binding energy shows appreciable differences. These results are
qualitatively comparable, although for high densities the standard
DDHFT treatment gives a softer growth for both $E_B$ and $P$.
These differences could be significative for those phenomena
dominated by the regime of extreme densities, such as the
structure of neutron stars.

\begin{figure}\vspace{-3cm}
\includegraphics[width=\textwidth]{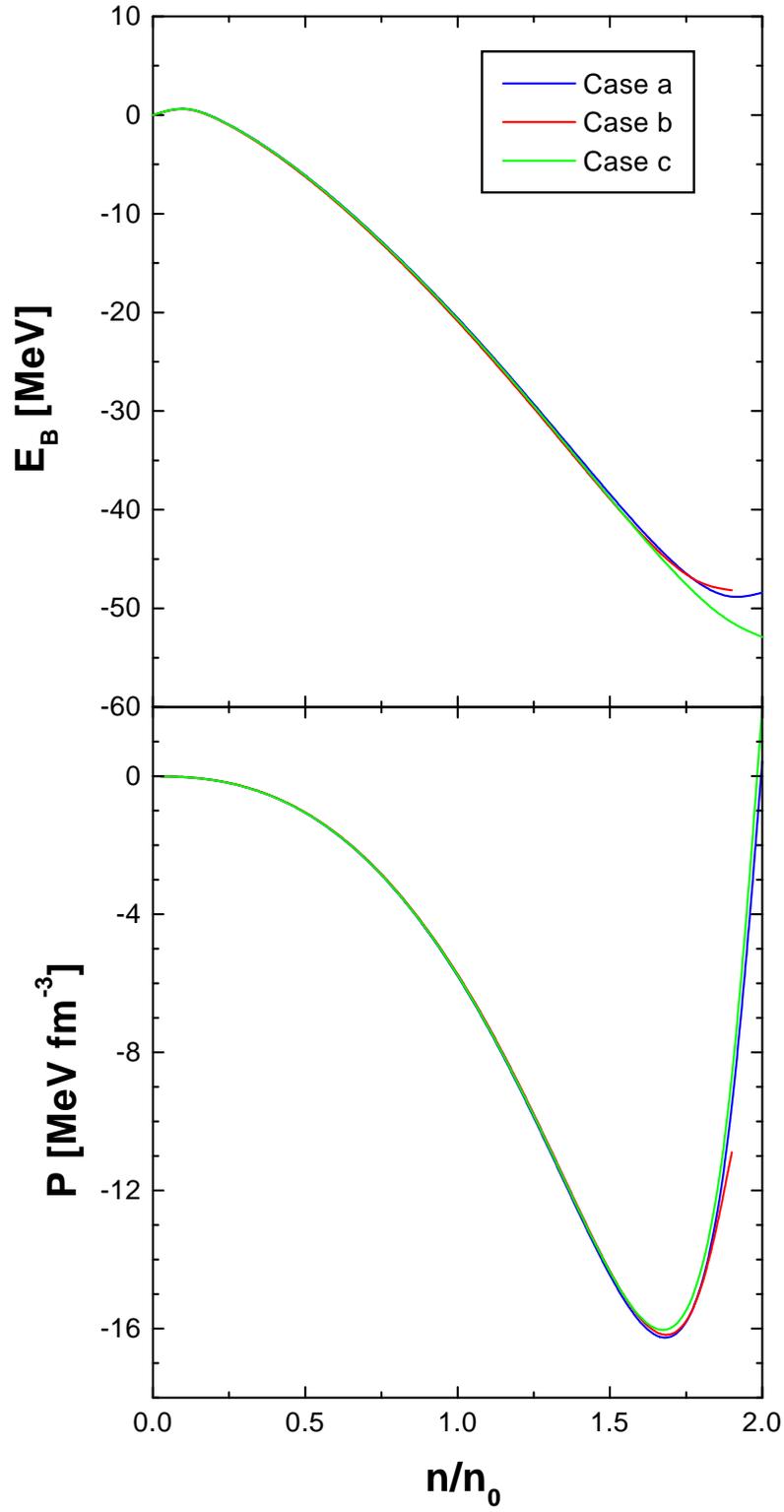}
\caption{The binding energy (top) and pressure (bottom) obtained
by using as input the results of \cite{DRUKAREV}, within the
approaches (a-c). \label{FIG2}}

\end{figure}

\begin{figure}\vspace{-3cm}
\includegraphics[width=\textwidth]{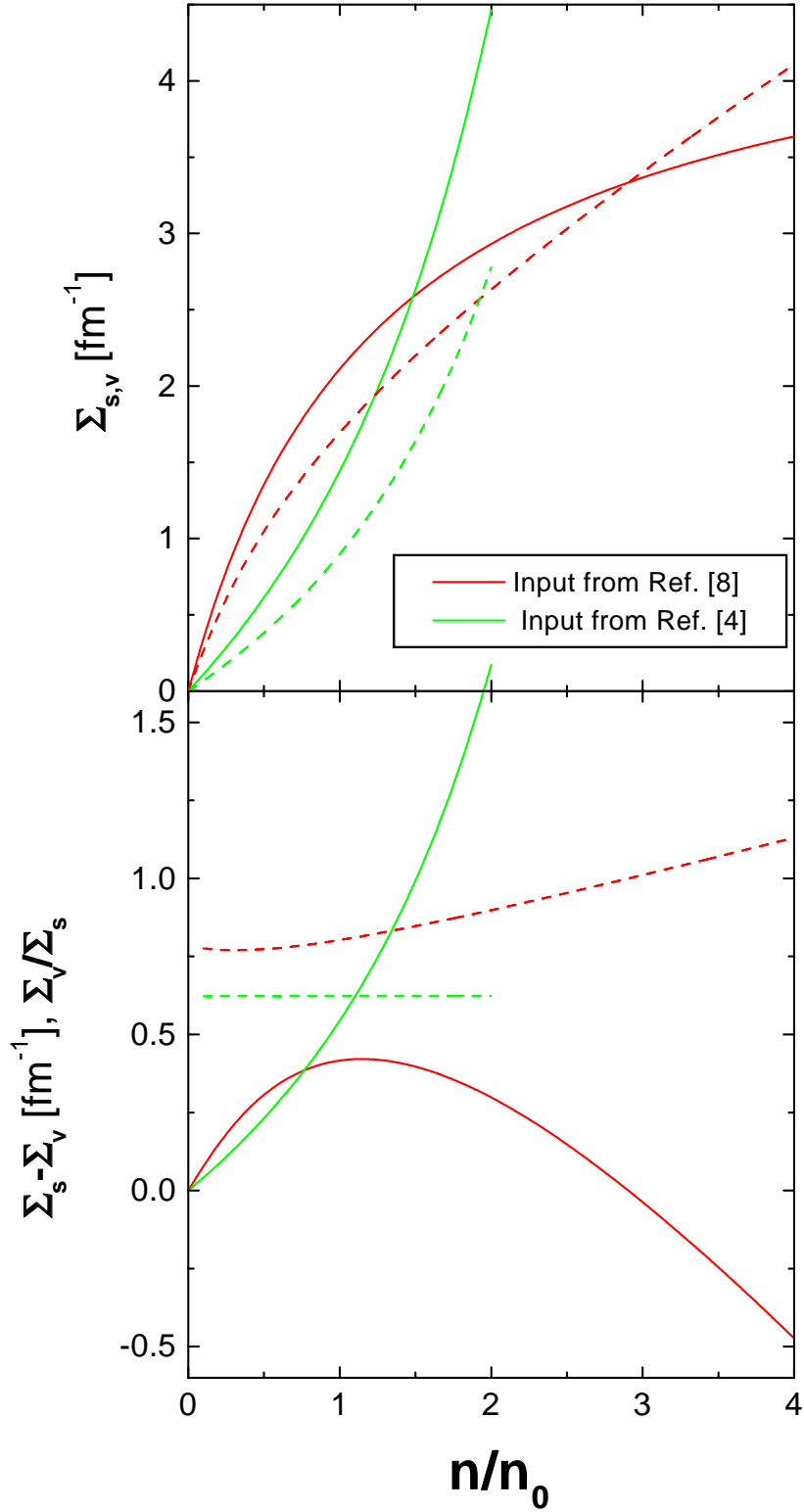}
\caption{The nucleon self-energies for symmetric nuclear matter,
provided by references \cite{TYPEL} and \cite{DRUKAREV}. In the
upper panel the magnitudes of $\Sigma_s^{(in)}$ (solid lines) and
$\Sigma_v^{(in)}$ (dashed lines), in the lower panel the
difference (solid lines) and the quotient (dashed lines) between
the scalar and vector components.\label{FIG3}}
\end{figure}

\begin{figure}\vspace{-3cm}
\includegraphics[width=\textwidth]{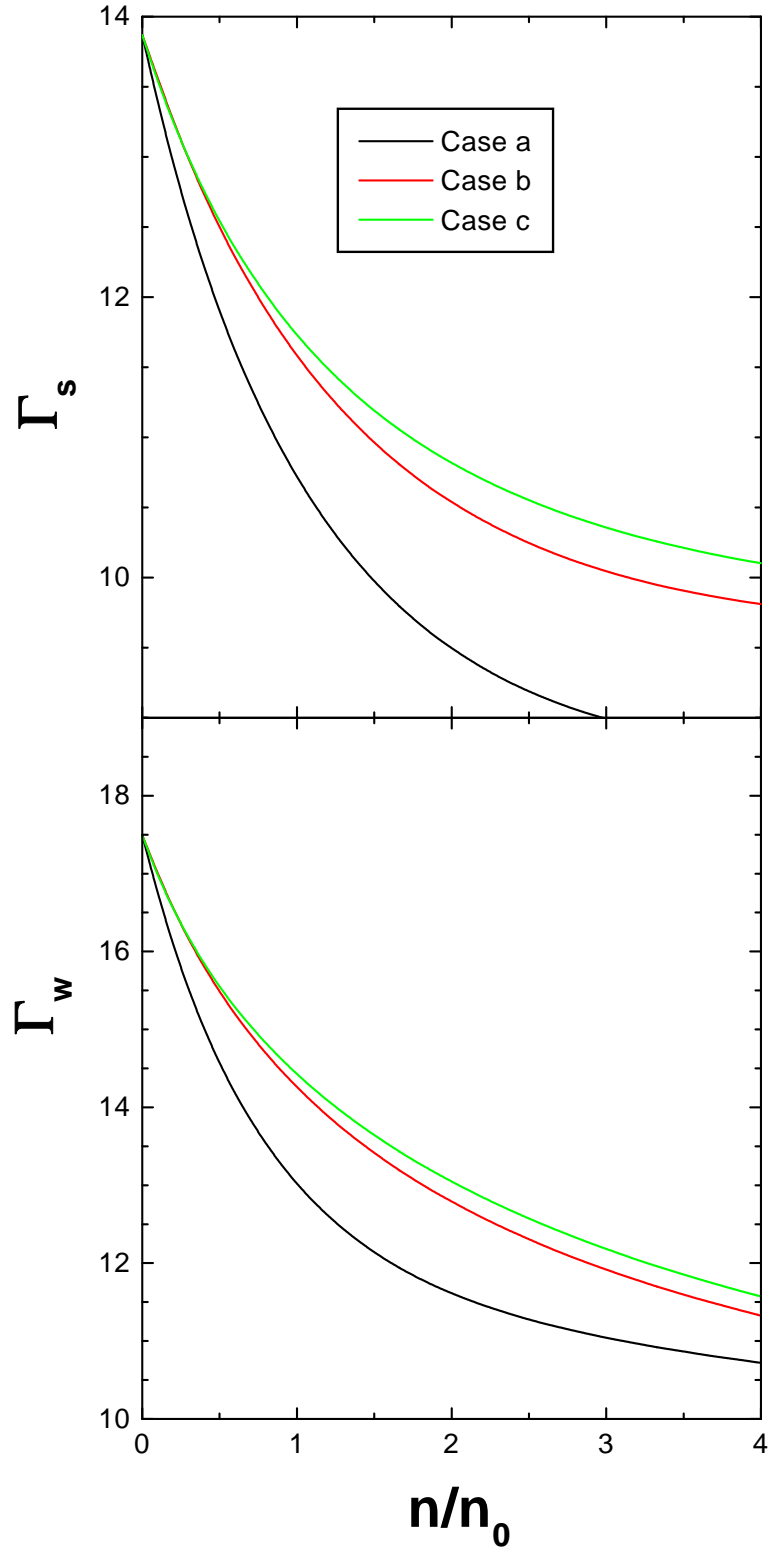}
\caption{The effective nucleon-meson vertices as functions of the
baryonic density, obtained by using the self-energy deduced from
\cite{TYPEL} under the cases (a-c). \label{FIG4}}
\end{figure}

As the next subject we have examined the effect of multiple
representations. Indeed, given a set of functions
$\Sigma_s,\,\Sigma_\mu$ depending on the macroscopic variables of
nuclear matter, there exist an indefinite number of models like
that of Eq. (\ref{LAGRANGIAN}), capable of reproduce them within
MFA. Some of these possibilities have been listed as cases (a-c)
in the previous section. The formal aspects have been summarized
in Eqs. (\ref{CASEA})-(\ref{CASEC}), now we focus on some
thermodynamical observables of symmetric nuclear matter.\\
To avoid biased conclusions we have considered two inputs of
diverse source. In addition to the results of \cite{TYPEL}, based
on one-boson exchange potentials, we have included the QCD sum
rules calculations of \cite{DRUKAREV}. We have evaluated the
binding energy and the pressure, for each of these inputs under
the three assumptions (a-c). We have found that despite the
different parameterizations used for a given input, the results
are practically undistinguishable in both cases. In Fig.
\ref{FIG2} we show this feature for the calculations using the
inputs obtained from \cite{DRUKAREV}, where it can be seen that
the three curves are nearly coincident. The range of densities has
been restricted to the region of validity of the QCD sum rules
computations. The binding energy is monotonically decreasing, it
seems to have a minimum near $n\simeq 2\,n_0$ where the effective
nucleon mass tends to zero. This description does not adjust to
the nuclear matter phenomenology, however we keep Ref.
\cite{DRUKAREV} into consideration as it offers a physically
meaningful input. It must be noted that the result (c) does not
coincide with the preliminary result given in Ref. \cite{AGUIRRE},
as the $\omega$ meson contribution to the energy was not properly
taken into
consideration there.\\
A similar conclusion is obtained by using the self-energies
extracted from \cite{TYPEL}, since the curves corresponding to the
instances (b) and (c) (not shown in the figure) follow closely the
solid curves of Fig. \ref{FIG1}, in the full range $0<n<4\,n_0$.\\
The different behavior between the inputs obtained from
\cite{DRUKAREV} and \cite{TYPEL}, for the energy per particle, can
be justified by examination of Fig. \ref{FIG3}. In the upper panel
the self-energies as functions of the density, show a monotonous
increase in both cases. However the rate of growth of $\Sigma_s$
becomes lesser than the corresponding to $\Sigma_v$ for densities
$n \geq n_0$, in the parameterization of \cite{TYPEL}. The QCD sum
rules outputs instead, keeps a faster increase of the scalar as
compared to the vector self-energy, over all the domain of
densities. This behavior is emphasized in the bottom panel, where
the difference $\Sigma_s-\Sigma_v$ is monotonous increasing in one
case, but exhibits a extremum in the other one. Furthermore, a
constant quotient $\Sigma_v/\Sigma_s$ is obtained for the QCD sum
rules input, but it increases slowly for the other instance.
Anyhow, the rational density dependence of Eq. (\ref{QCDSRself})
for the iso-scalar self-energies could be retained in order to
simulate the nuclear matter mechanism of saturation, for instance
with $\alpha_s=-0.278, \,\alpha_v=-2.93,$ and $\lambda=0.697$ we
have obtained a binding energy of $-14$ MeV at $n\simeq 1.1\,n_0$.
These values must be contrasted with the parameterization deduced
from \cite{DRUKAREV}, $\alpha_s=-2.309, \,\alpha_v=-0.66,$
and $\lambda=-0.261$.\\
The density dependence of the vertices, obtained by using the
input of Ref. \cite{TYPEL}  under the three approaches (a-c) is
presented in Fig. \ref{FIG4}. It must be noted that the
expressions for $\Gamma_{s,w}$ become singular at zero density,
nevertheless they have finite limits: $\Gamma_s^2\rightarrow m_s^2
d\Sigma_s^{(in)}/dn, \,\Gamma_w^2\rightarrow m_w^2
d\Sigma_v^{(in)}(0)/dn$, for all the
cases under consideration.\\
A common feature of the vertices is a strong decrease up to $n
\leq 2\, n_0$, which moderates above this value. The cases (b) and
(c) yield very similar results for both $\Gamma_s$ and $\Gamma_w$,
differences with the case (a) may lead to qualitatively distinct
descriptions for subsequent applications. All these results for
the vertices $\Gamma_k(z)$ can be summarized by an expression in
terms of its natural variable $z$,
$\Gamma_k=(z+a_k)/\left(b_k+c_k(z+a)+d_k(z+a)^2\right)$, with
$k=s,w$ and the numerical coefficients are shown in Table 1.

\begin{table}
\begin{tabular}{|lcccccccccc|}
\hline Case&$a_s$&$b_s$&$c_s$&$d_s$&$a_w$&$b_w$&$c_w$&$d_w$&$z_s$&$z_w$\\
\hline
(a)&  0.805&-0.170&0.350&-0.082&0.386&-0.032&0.148&-0.024&$s_1$&$s_1$\\
(b)&  1.264&-0.232&0.294&-0.031&0.413&-0.018&0.096&0.009&$s_2$&$s_1$\\
(c)&  0.059&0.000&0.066&0.078&1.178&-0.217&0.285&-0.037&$m_2$&$m_1$\\
\hline
\end{tabular}
 \caption{Numerical values of the fitting coefficients for the
  density dependence of the vertices,
obtained in the approaches (a-c) using the self-energies deduced
from \cite{TYPEL}.}
\end{table}

Another insight into the basic phenomenology of nuclear matter,
can be given by the Landau parameters. They can be evaluated as
the Legendre projections of the second derivative of $E_{MFA}$
with respect to the baryonic density, for a deduction see
\cite{MATSUI}. These parameters are very useful in regard to
collective phenomena of the dense nuclear environment, as for
instance, phase transitions, the giant monopolar and quadrupolar
modes, or the sound  velocity \cite{KURASAWA}. In order to keep
track of the momentum dependence, we assume a non-zero spatial
component of the self-energy $\vec{\Sigma}_v$, the baryonic
current $\vec{j}$ and the omega meson $\vec{\omega}$, taking the
zero limit of these quantities at the end of the calculations. So,
for example, we write $E_p=\sqrt{(\vec{p}-\vec{\Sigma})^2+M^{\ast
\,2}}$ into the integral of Eq. (\ref{EOS}). Denoting by $n_k$ the
occupation number of nucleon states with a well defined momentum
$\vec{p}_k$, then we define
\begin{eqnarray}
f_{kl}&=&\frac{\partial^2 E_{MFA}}{\partial n_k \partial
n_l} \\
&=&-\frac{M^\ast}{E_i}\,\frac{\partial\Sigma_s}{\partial
n_l}+\frac{\partial\Sigma_v}{\partial
n_l}-\frac{\vec{p}_k}{E_k}\cdot \frac{\partial
\vec{\Sigma}_v}{\partial
n_l}-\frac{\partial\vec{\Sigma}_v}{\partial
n_k}\cdot\left(\frac{\partial \vec{j}}{\partial
n_l}-\frac{n_s}{M^\ast} \frac{\partial\vec{\Sigma}_v}{\partial
n_l}\right)\nonumber +m_w^2\frac{\partial \vec{\omega}}{\partial
n_k}\cdot\frac{\partial \vec{\omega}}{\partial n_l}, \nonumber
\end{eqnarray}

in the second equality, the limit of isotropic nuclear matter has
been taken. The derivative of the baryonic current can be
evaluated in the MFA
\[
\frac{\partial \vec{j}}{\partial
n_i}=\frac{\vec{p}_i}{E_i}-H\,\frac{\partial
\vec{\Sigma}_v}{\partial n_i},\;\;\;
H=\frac{1}{6\pi^3}\int_0^{p_f}d^3p\;\frac{2p^2+3 M^{\ast
\,2}}{(p^2+M^{\ast \,2})^{3/2}}.\nonumber \]

\begin{figure}
\vspace{-3cm}
\includegraphics[width=\textwidth]{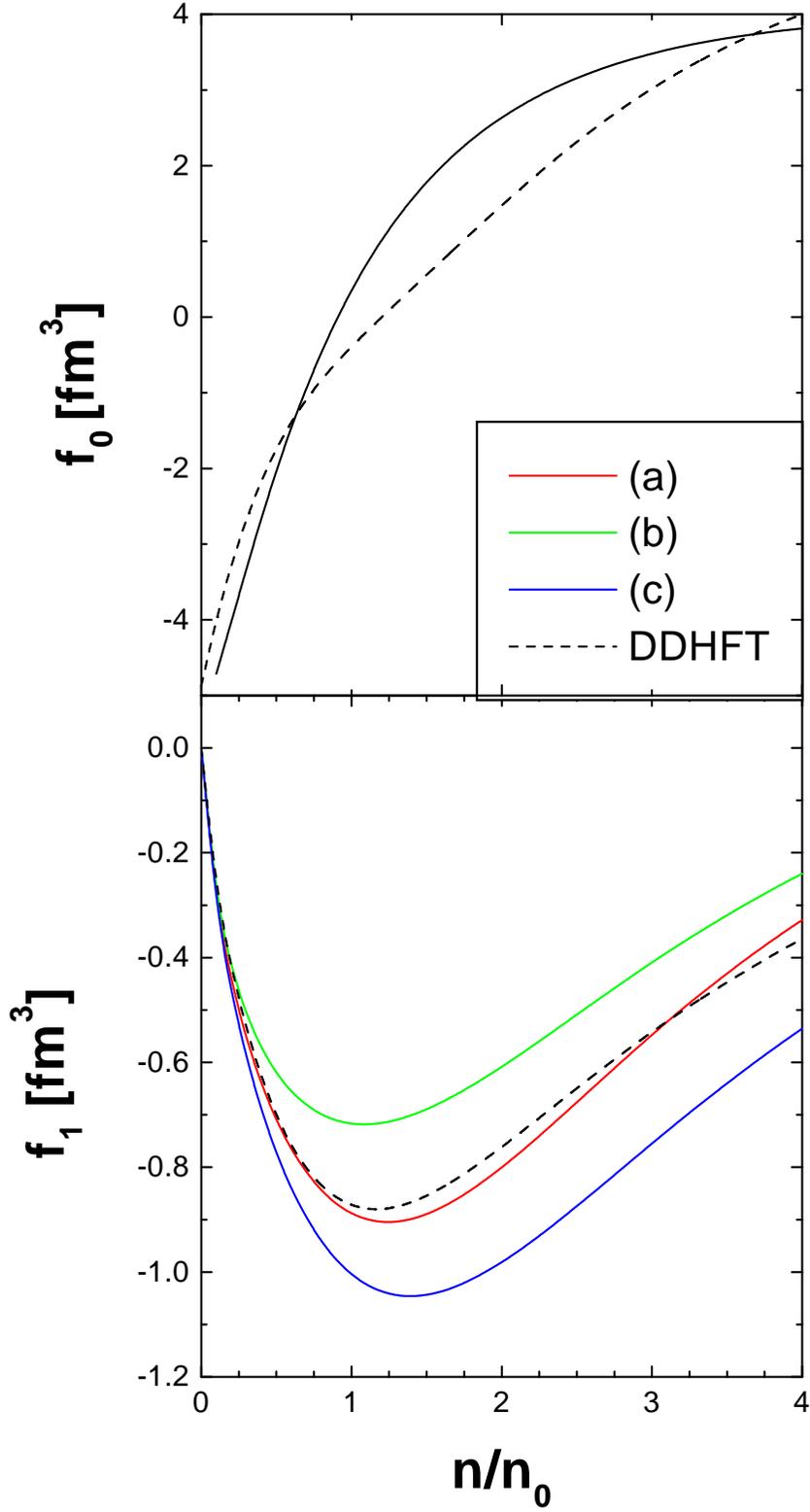}
\caption{The Landau parameters of zero and first order in terms of
the baryonic density, obtained by using the self-energy deduced
from \cite{TYPEL}, and under the cases (a-c) and the standard
DDHFT treatment.\label{FIG5}}
\end{figure}

\begin{figure}
\vspace{-3cm}
\includegraphics[width=\textwidth]{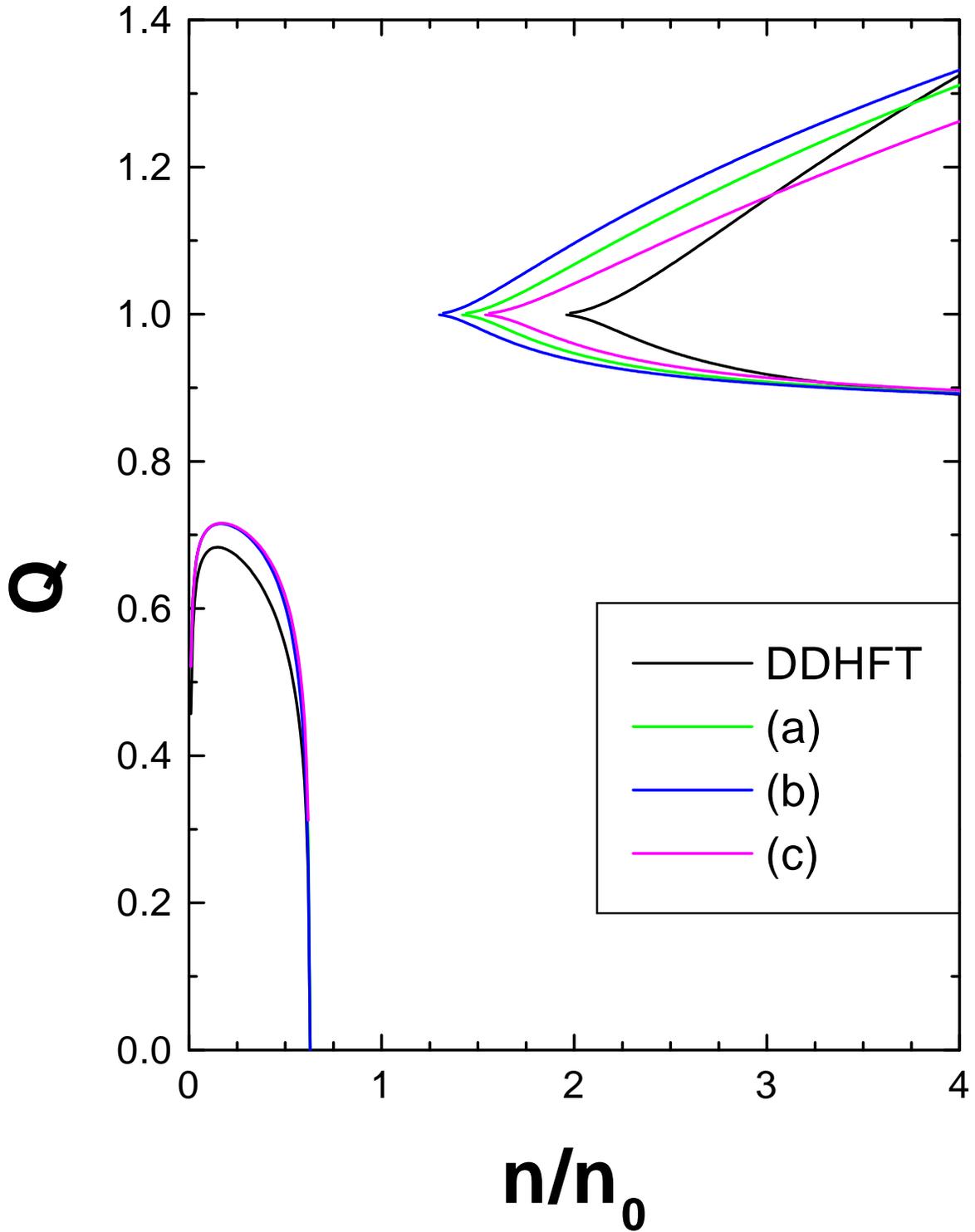}
\caption{The low lying collective longitudinal modes in terms of
the baryonic density, under the cases (a-c) and the standard DDHFT
treatment.\label{FIG6}}
\end{figure}

The remaining derivatives are
\[
\frac{\partial \vec{\omega}}{\partial
n_i}=\frac{\tilde{\omega}}{n}\,\frac{\partial \vec{j}}{\partial
n_i}, \;\;\; \frac{\partial \vec{\Sigma}_v}{\partial
n_i}=\frac{\vec{p}_i}{E_i}\,\frac{\Sigma_v}{n+H\,\Sigma_v},
\]
whose explicit form depends on the approach used (a-c). \\The
Landau parameters $f_k$ are defined by means of the projections
\[
f_m=\frac{2m+1}{2\pi}\int_{-1}^{1}f_{kl} P_m(\nu) d\nu,
\]
with $\nu$ the cosine of the angle between $\vec{p}_k$ and
$\vec{p}_l$, and $P_m$ the Legendre polynomial of order $m$. At
the end of the calculations, the replacement $\mid\vec{p}_k\mid,
\,\mid\vec{p}_l\mid\;\rightarrow p_F$ is made. By this procedure,
we have obtained the Landau parameters of zero and first orders
\begin{eqnarray}
f_0&=&-\frac{M^\ast}{E_F}\,\frac{d\Sigma_s^{(inp)}}{dn}+\frac{d\Sigma_v^{(inp)}}{dn},
\nonumber \\
f_1&=&\left(\frac{p_F/E_F}{n+H\,\Sigma_v^{(inp)}} \right)^2\left[
m_w^2 \tilde{\omega}^2-2 n
\Sigma_v^{(inp)}+\left(\frac{n_s}{M^\ast}-H
\right)\Sigma_v^{(inp)\,2}\right]. \nonumber
\end{eqnarray}
It is costumary to normalize the Landau parameters with the
density of states at the Fermi surface $\varepsilon_F=2 p_F
E_F/\pi^2$, so that $F_k=\varepsilon_F f_k$.\\
For a given model of self-energies, the result for $f_0$ does not
depend on the field parameterization of the vertices. The
parameter $f_1$ instead, depends on $\tilde{\omega}$ which
certainly differs for each of the instances (a-c). In Fig.
\ref{FIG5} the density dependence of both Landau parameters is
shown, significative departures for $f_1$ are found for densities
above $0.25 n_0$.\\
Consequently all physical magnitudes depending solely on $f_0$,
such as the nuclear compressibility $K=3p_F^2(1+F_0)/E_F$, do not
distinguish among the field parameterization of the vertices. The
same feature is shared by the giant monopole mode energy
$E_M=\alpha_0/A^{1/3}$, with $\alpha_0\simeq 1.076 \sqrt{K/\mu}$,
and the first sound velocity $v_1=p_F\sqrt{(1+F_0)/(3\mu E_F)}$.
On the other hand, the excitation energy of the quadrupole state
is given by $E_Q=\alpha_2/A^{1/3}$, with
$\alpha_2=p_F\sqrt{2}/(1.2\mu\sqrt{1+F_1/3})$, therefore it is
sensitive to the approach used. We have obtained in our
calculations $K\simeq440$ MeV, $\alpha_0\simeq145$ MeV, and
$\alpha_2\simeq77$ MeV at the normal density, which must be
compared with the empirical values 230 MeV $<K<$ 270 MeV,
$\omega_0=80$ MeV, and $\omega_2=63$ MeV.\\
Another interesting phenomenological property that can be related
to the Landau parameters, are the zero sound modes. These are
longitudinal collective modes propagating in nuclear matter at
zero temperature, whose dispersion relation can be found as the
zeros of the longitudinal dielectric function
\[
\varepsilon_L=1+\left( F_0+\frac{F_1\,Q^2}{1+F_1/3}\right)\Phi(Q)
\]
with $Q=p_0 E_F/(\mid \vec{p}\mid p_F)$ and $\Phi$ the Lindhard
function. Collective modes have been studied in detail within
relativistic field models, see for instance
\cite{MATSUI,KURASAWA,LIM,CAILLON,COLLMODES}. Low lying collective
modes can be classified as instability modes, in the low density
regime, and zero sound modes with a characteristic linear
dispersion relation $p_0\propto \mid \vec{p}\mid$. In Fig.
\ref{FIG6}, the dispersion relation in terms of the baryonic
density is shown for the standard DDHFT treatment and for the
approaches (a-c), within the self-energy parametrization of
\cite{TYPEL}. These results are appropriate for the regime $\mid
\vec{p} \mid \rightarrow 0$, but keeping the linear dispersion
relation. The instability and zero sound modes are clearly
distinguishable, as the first one is closely related to the
equation of state, all the corresponding curves practically
coincide. The zero sound yields very different behaviors, the
extremum values for its threshold are obtained within
 the approaches (b) and the standard DDHFT treatment. We have obtained $n/n_0=1.25$
and $2$, respectively. In all the cases we have obtained
two-folded zero sound, and taking into account that $\Phi(Q)$ has
non zero imaginary part for $Q<1$, only the upper mode is
undamped.

\section{Conclusions}\label{SecIV}
We have examined in this work, two definitions for the vertices of
a hadronic field model in terms of the self-energies provided by
another theoretical framework. The first one is the standard DDHFT
treatment, which uses Eqs. (\ref{DDHFT}) to define the vertices
$\Gamma_s,\, \Gamma_w$ between nucleon and scalar and vector
mesons, respectively. There, $\Sigma^{(in)}$ denotes the nucleon
self-energies, based on Dirac-Brueckner calculations with
one-meson exchange potentials. Although vector and scalar
dependencies have been stated \cite{LENSKE}, most practical
applications have used only the first one, similar to our case
(a). This particular choice has the property of reproducing the
equation of state of the source calculations, but on the contrary
the self-energies formally differ from the inputs $\Sigma^{(in)}$.
An alternative definition was proposed in \cite{AGUIRRE}, it
consisted in equating both, the input self-energy and that
evaluated within the hadronic model with vertices depending on
scalar combinations of hadronic fields. This procedure ensures a
coherent overlap of the effective hadronic field model and a more
\textit{fundamental} theoretical description, at least in the
baryon sector. Three different parametrizations of the vertices,
cases (a-c), and two theoretical sources, namely the QCD sum rules
of Ref. \cite{DRUKAREV} and the Dirac-Brueckner parametrization of
Ref. \cite{TYPEL}, have been
examined within this scheme.\\
Thermodynamical aspects of symmetric nuclear matter evaluated in
MFA, are undistinguishable for the instances (a-c) within the
approach of \cite{AGUIRRE}, see Eqs.(\ref{CASEA}-\ref{CASEC}), but
they differ noticeably from the standard DDHFT results. Although
the almost perfect coincidence within the (a-c) approaches for the
energy and pressure, the vertices obtained present significative
deviations in the medium to high densities regime. The results for
$\Gamma_s, \, \Gamma_w$ are similar in the (b) and (c) instances,
as they use parametrizations in terms of variables strongly
related in the MFA. We have given fittings for all the
vertices in terms of its corresponding variables.\\
In reference to the theoretical source for the inputs
$\Sigma^{(in)}$, the parameterization given in \cite{TYPEL} yields
observables in agreement with the nuclear matter phenomenology, as
expected. The QCD sum rules instead, do not provide an acceptable
description, at least in its present status. Taking the nucleon
self-energies proposed in \cite{DRUKAREV}, we have presented
formal expresions for the vertices of a hadronic field model for
both isoscalar and isovector channels. However, examination of the
binding energy obtained for symmetric nuclear matter, show a
qualitative mismatching. This assertion is in opposition to the
preliminary results of \cite{AGUIRRE}, where the vector meson
contribution to the energy density was not taken into account
correctly. However we have shown that the functional dependence of
the isoscalar self-energies given in \cite{DRUKAREV} is not
completely reasonless, so as a redefinition of its numerical
parameters produce a satisfactory equation of state.\\
We have also considered in detail the Landau parameters for
symmetric nuclear matter, under the four assumptions and the
parametrization of \cite{TYPEL}. The hadronic model yields two
non-zero Landau parameters $F_0$ and $F_1$, the first one has a
common behavior while the second one strongly depends on the
vertices definition. Therefore, nuclear observables depending on
$F_0$ only, such as isothermal compressibility, first sound
velocity and the giant monopole energy, give almost coincident
values. Those quantities depending on $F_1$ may discriminate among
the different approaches, for instance the zero sound modes
exhibit very different threshold densities according to the choice
made.\\
To sum up, we have compared several formal schemes of merging
external physical information into hadronic field models. For this
purpose we have introduced interaction vertices $\Gamma_s$ and
$\Gamma_w$, which are functionals of the hadronic fields, and we
have determined them by else the standard DDHFT procedure or by
requiring the reproduction of the self-energies used as input.
Either of them, yield comparable results for the binding energy
and pressure of symmetric nuclear matter. However, physical
quantities can be found able to distinguish among the different
approaches.\\
QCD sum rules calculations for the nucleon self-energy, do not
meet the requirements of the procedure, however further
refinements could improve its performance.

\newpage

\newpage

\newpage

\newpage

\newpage

\newpage
\end{document}